\documentstyle[12pt]{article}
\begin{document}
\oddsidemargin .4cm
\begin{titlepage}
\title{\hfill {\normalsize IFT/13/94}\\
\hfill{\normalsize hep-ph/9412236}\\
\hfill{\normalsize revised version}\\
\vspace{1.5cm}
 Improved analysis of the renormalization\\
 scheme ambiguities in the QCD corrections\\
 to the semileptonic decay of the tau lepton\thanks{Supported by
 the KBN grant 202739101}}
\vspace{3.cm}
\author{ Piotr A. R\c{a}czka\thanks{E-mail: praczka@fuw.edu.pl}
 \hspace{.3cm}
 and  Andrzej Szymacha\\
 Institute of Theoretical Physics\\
 Department of Physics, Warsaw University\\
 ul.\ Ho\.{z}a 69, 00-681 Warsaw, Poland.}
\date{\quad}
\maketitle
\begin{abstract}
\noindent
 The perturbative QCD corrections
 to the semileptonic decay width of the tau lepton
  are evaluated in the next-next-to-leading order from the
 contour integral representation in various renormalization
  schemes, using numerical solution
 of the renormalization group equation for complex energies.
 A  quantitative estimate of the
 ambiguities resulting from the freedom of choice of the
 renormalization scheme is obtained by taking into account predictions
 in all schemes that do not involve large cancellations in the
 expression for the scheme invariant combination of the expansion
 coefficients.
 The problem of an optimal choice of
 the renormalization scheme for the improved perturbative
 expression is discussed.  A fit of $\Lambda_{\overline{MS}}$
 is made using the available  experimental data.
\end{abstract}
\thispagestyle{empty}
\end{titlepage}
\setcounter{page}{1}
\newpage
 As is well known, predictions of a quantum field theory obtained with
 finite order perturbative expressions depend on the choice of the
 renormalization scheme (RS). This effect is usually regarded as a
 complication in the comparison of the theory with the experimental data.
 However, the effect of renormalization scheme dependence may be also used
 to improve our understanding of the perturbative result. By estimating the
 change in the predictions over a set of {\em a priori}
 acceptable schemes we may test
 reliability of the perturbative expression. This point of view has been
 recently discussed in \cite{b,c}, where a
 concrete method has been proposed for
 a systematic evaluation of the RS ambiguities in the QCD predictions in
 the next-next-to-leading order (NNLO). This method is based on the
 observation, that one should take into account predictions in all
 schemes that do not involve large cancellations in the expression for
 scheme independent combination of the expansion coefficients. Application
 of this method was illustrated in \cite{c} using as an example the QCD
 corrections to the Bjorken sum rule for the polarized structure
 functions. Also the problem of the RS ambiguities in the QCD corrections
 to the ratio
 \cite{lamyan}-\cite{branapi}:
\begin{equation}
R_{\tau}=\frac{\Gamma(\tau \rightarrow \nu_{\tau} + hadrons)}
{\Gamma(\tau \rightarrow \nu_{\tau} e^{-} \overline{\nu}_{e})},
\end{equation}
 has been considered in this framework \cite{b}. In this case it was found,
 that even small changes in the RS - among acceptable schemes - resulted
 in a large variation in the predictions. The strong RS dependence seemed to
 undermine the phenomenological importance of the QCD corrections to
 $R_{\tau}$.

 However, the  QCD correction to $R_{\tau}$ comes
 originally in the form of a contour integral in the complex energy
 plane \cite{lamyan}-\cite{bra89}.
 The commonly used perturbative formula appears as
 a result of approximate evaluation of the contour integral.
 As was pointed out
 in \cite{piv,ledepi}, the evaluation of the contour integral may
 be improved. In \cite{ledepi} it was shown that
 using the improved evaluation of the contour integral one obtains
 smaller scale dependence in the ${\overline{MS}}$ scheme.
 It is then an interesting problem,
 how the improved accuracy of the predictions affects the estimate
 of the strength of the RS dependence evaluated according to the
 method developed in \cite{b,c}.

 In this note we evaluate the QCD corrections to $R_{\tau}$
 in various renormalization schemes by solving
 numerically the renormalization group equation for complex energies,
 and by computing numerically the relevant contour integral.
 Using
 the method developed in \cite{b,c} we obtain a quantitative estimate
 of the strength of the RS dependence in this case \cite{rszym}.
 We discuss the
 problem of an optimal choice of the renormalization scheme in
 the improved expression. We also make a
 fit of $\Lambda_{\overline{MS}}$ taking into account
 recently obtained  experimental results for $R_{\tau}$.

 Let us begin with a brief summary of the ideas
 presented in \cite{b,c}.   Let us consider
 a NNLO expression for a physical quantity $\delta$,
 properly normalized,  depending
 on a single energy variable $P$:
\begin{equation}
\delta(P^{2}) = a(P^{2})[1+r_{1}a(P^{2})+r_{2}a^{2}(P^{2})],
\label{R}
\end{equation}
 where $a(\mu^{2})=g^{2}(\mu^{2})/(4 \pi^{2})$
 denotes the running coupling constant that satisfies the NNLO
 renormalization group equation:
\begin{equation}
\mu \frac{da}{d\mu} = - b\,a^{2}\,(1 + c_{1}a + c_{2}a^{2}\,),
\label{rge}
\end{equation}
 (The effects of the quark masses are neglected.)
The
 value of the predictions obtained for $\delta$ with the NNLO
 expressions (\ref{R}) and (\ref{rge}) depends on the choice of RS.
 Presently in the
 phenomenological applications the modified minimal
 subtraction scheme ($\overline{MS}$)  \cite{msb} is used.
 However, other choices of the renormalization scheme are
 possible, all related to the $\overline{MS}$ scheme by a finite
 renormalization. In the NNLO, in the class of the mass and
 gauge independent schemes, the freedom of choice of the scheme
 may be characterized by two continuous parameters. (The relevant
 formulas describing the RS dependence of the coefficients
 $r_{1}$, $r_{2}$ and $c_{2}$ have been collected in \cite{b}.)
 The consistency of the perturbation theory guarantees,
  of course, that differences in the predictions obtained in
 various  schemes  are formally of higher
 order in the coupling constant, but numerical magnitude
 of these differences
 is significant for phenomenological analysis. This fact stimulated
 the interest in prescriptions for making a proper choice of
 the RS
 \cite{CelGon}-\cite{DukeRoberts}.
 The most attractive of these propositions seems to be the choice
 based on the so called Principle of Minimal Sensitivity (PMS)
 \cite{PMS}.

However, whatever scheme we decide to
 choose as an optimal one, there is a continuum of equally
 reasonable schemes close to the one preferred by us.
 Predictions
 in such schemes also should be somehow taken into account
 in the phenomenological analysis. A natural way to do this is
 to supplement the prediction in a preferred scheme with an
 estimate of the variation of the predictions over the whole set
 of {\em a priori} acceptable schemes.  A concrete
 realization of this idea was presented in \cite{c}, based on the
 existence of an  RS independent combination of the
 expansion coefficients \cite{PMS,Dhar,DharGupta,GrunEC}:
\begin{equation}
\rho_{2}=c_{2}+r_{2}-c_{1}r_{1}-r_{1}^{2}.
\label{rho2}
\end{equation}
 which provides a natural RS
 independent characterization of the magnitude of the
 NNLO corrections for the considered physical quantity.
 In \cite{c} it was
 proposed to  calculate variation of the predictions for $\delta$ over the
 set of schemes for which the expansion coefficients
 satisfy the condition:
\begin{equation}
\sigma_{2}(r_{1},r_{2},c_{2}) \leq l\,|\rho_{2}|,
\label{constraint}
\end{equation}
 where
\begin{equation}
\sigma_{2}(r_{1},r_{2},c_{2})=|c_{2}|+|r_{2}|+c_{1}|r_{1}|+r_{1}^{2}.
\end{equation}
  A  motivation for the condition (\ref{constraint}) is
 that it eliminates schemes in which the expressions (\ref{R})
 and (\ref{rge}) involve unnaturally large  expansion
 coefficients, implying large cancellations in the expression
 for the RS invariant $\rho_{2}$. The constant $l\,$ in the
 condition (\ref{constraint}) controls the degree of cancellations
 that we want to allow in the expression for $\rho_{2}$.
 As was pointed out in \cite{c}, taking $l=2$ in (\ref{constraint}) one
 has the PMS scheme right at the boundary of the allowed region
 in the space of parameters characterizing the scheme.

 It is expected that the estimate of the strength of the
 RS dependence obtained according to this prescription would be
 useful for a quantitative comparison of reliability of perturbative
 predictions for different physical quantities, evaluated
 at different energies. It should be also very useful in
 determining the regions of applicability of the perturbative
 expansion. Indeed, any large variation of the predictions
 over a set of schemes satisfying constraint (\ref{constraint}) with
 $l=2$ would be an unambiguous sign of a limited applicability of
 the NNLO expression.

 Let us now consider the
 QCD corrections to the semileptonic decay
 of the tau lepton. These corrections, denoted further by
 $\delta_{\tau}$,  contribute to the ratio $R_{\tau}$ in the
 following way:
\begin{equation}
R_{\tau}=3(|V_{ud}|^{2}+|V_{us}|^{2})S_{EW}
(1+\tilde{\delta}_{EW}+\delta_{\tau})
\label{rt}
\end{equation}
 where  $V_{ij}$ are the Cabbibo-Kobayashi-Maskawa matrix elements
 and $S_{EW}=1.0194$,  $\tilde{\delta}_{EW}=0.001$ are factors arising from
 the electroweak corrections \cite{marsir,brali}.
  The perturbative correction to $\delta_{\tau}$ is known up to NNLO
 \cite{bra88}-\cite{kataev} in the approximation
 of massless {\em u}, {\em d} and {\em s} quarks, and it has
 in the conventional approach  the form
 (\ref{R}) with the characteristic energy scale $P=m_{\tau}=1.7771$~GeV.

 The interest in the QCD corrections to $R_{\tau}$
 \cite{bra88}-\cite{ledepi}, \cite{maxnich}-\cite{katstar}
  was  originally stimulated by the fact,
 that the perturbative expression for
 $\delta_{\tau}$ appeared to be quite sensitive to
 the QCD parameter $\Lambda$.
 Thus in principle a very good test of perturbative
 QCD predictions could be obtained \cite{bra88,narpi},
 since the effects of nonzero masses of light quarks,  and
 the nonperturbative effects, were estimated to be small
 despite the rather low energy scale.
 Unfortunately, in \cite{b}
 large differences in the perturbative predictions for  $\delta_{\tau}$
 have between obtained in {\em a priori}
 acceptable schemes. (The perturbative
 QCD corrections to $R_{\tau}$ in some optimal schemes
 were considered previously in
 \cite{maxnich,haru90,chylkat,haru92}.)
  It seemed therefore that the
 high  sensitivity of $\delta_{\tau}$ to the QCD parameter $\Lambda$ is
 practically eliminated by the ambiguities introduced by the strong RS
 dependence \cite{pr95}.

However, the commonly used perturbation
expansion for $\delta_{\tau}$ involves not only the obvious approximation
resulting from the truncation of the perturbation series, but also an
approximation of another kind, which appears in the process of
evaluation of $\delta_{\tau}$ from the expression for the two-point quark
current correlators. Indeed,
in order to evaluate the QCD corrections to $R_{\tau}$ one starts
from the formula \cite{lamyan,schil,bra88,bra89}:
\begin{equation}
R_{\tau}=12\pi\int_{0}^{m_{\tau}^{2}}\frac{ds}{m_{\tau}^{2}}
\left(1-\frac{s}{m_{\tau}^{2}}\right)^{2}
\left[\left(1+2\frac{s}{m_{\tau}^{2}}\right)
Im\Pi^{(1)}(s+i\epsilon)+...\right],
\label{imp}
\end{equation}
where $\Pi^{(1)}$ denotes the sum of the transverse parts of the
$\Delta\,S=0,1$ vector and axial quark current correlators:
\begin{equation}
\Pi^{(1)}(s)=|V_{ud}|^{2}[\Pi^{(1)}_{ud,V}(s)+\Pi^{(1)}_{ud,A}(s)]+
|V_{us}|^{2}[\Pi^{(1)}_{us,V}(s)+\Pi^{(1)}_{us,A}(s)],
\end{equation}
\begin{equation}
\Pi^{\mu\nu}_{kl,V/A}(q)=(-g^{\mu\nu}q^{2}+q^{\mu}q^{\nu})
\Pi^{(1)}_{kl,V/A}(q^{2})+...,
\end{equation}
\begin{equation}
\Pi^{\mu\nu}_{kl,V/A}(q)=i\int\,d^{4}xe^{iqx}<0|T(J^{\mu}_{kl,V/A}(x)
J^{\nu}_{kl,V/A}(0)^{\dagger})|0>.
\end{equation}
(If the quark mass effects are neglected, the longitudinal part of
$\Pi^{\mu\nu}$ does not contribute.  Also, the electroweak
contributions have been neglected.)  Since the QCD predictions for
the quark current correlators are not very well known for real
positive $s$, one uses the analyticity properties of $\Pi$ to convert
the expression (\ref{imp}) into the contour integral in the complex
energy plane:
\begin{equation}
R_{\tau}=\frac{6\pi}{i}\int_{C}\frac{ds}{m_{\tau}^{2}}
\left(1-\frac{s}{m_{\tau}^{2}}\right)^{2}
\left[\left(1+2\frac{s}{m_{\tau}^{2}}\right)\Pi^{(1)}(s)+...\right],
\end{equation}
 where $C\,$ is a contour running clockwise from
$s=m^{2}_{\tau}-i{\epsilon}$ to $s=m^{2}_{\tau}+i{\epsilon}$ and
avoiding the region of small $|s|$, i.\ e.\ on $C$ we have
$|s|=m_{\tau}^{2}$.  Integrating by parts along the contour,
one obtains:
\begin{equation}
\delta_{\tau}=\frac{1}{2\pi}\int_{-\pi}^{\pi}d\theta
\left(1+2e^{i\theta}-2e^{3i\theta}-e^{4i\theta}\right)
\left[\delta_{\Pi}(-s)|_{s=-m_{\tau}^{2}e^{i\theta}}\right],
\label{rtcont}
\end{equation}
where $\delta_{\Pi}(-s)$ is defined via the relation:
\begin{equation}
(-12\pi^{2})s\frac{d\,}{ds}\Pi_{V}^{(1)}(s)=
3(|V_{ud}|^{2}+|V_{us}|^{2})[1+\delta_{\Pi}(-s)].
\end{equation}
 The importance of this expression lies in the fact, that $\delta_{\tau}$
has been expressed via the quantity $\delta_{\Pi}(-s)$, which is formally RS
independent, and which is directly calculable in perturbative QCD for
real negative $s$. $\delta_{\Pi}(-s)$ has a perturbation expansion of the
form (\ref{R}), with the expansion coefficients in the $\overline{MS}$
scheme, with $\mu^{2}=-s$, given by \cite{kataev}:
\begin{equation}
r_{1}^{\overline{MS}}=1.63982, \,\,\,\,\,\, r_{2}^{\overline{MS}}=
6.37101.
\end{equation}
 The commonly used expansion of $\delta_{\tau}$ in terms of
$a(m_{\tau}^{2})$ is obtained from the expression (\ref{rtcont}) by
expanding $a(-s)$ under the integral in terms of $a(m^{2}_{\tau})$:
\begin{eqnarray}
\lefteqn{
a(-s)=a(m_{\tau}^{2})\left[1-
\frac{b}{2}\ln\left(-s/m_{\tau}^{2}\right)
a(m_{\tau}^{2})+\right.
}\nonumber\\
& & +\left.
\left(\left(\frac{b}{2}\ln\left(-s/m_{\tau}^{2}\right)\right)^{2}-
\frac{b}{2}\ln\left(-s/m_{\tau}^{2}\right)\right)
a^{2}(m_{\tau}^{2})\right],
\end{eqnarray}
 and
performing the contour integral over the appearing powers of
$\ln(-s/m_{\tau}^{2})$ explicitly.  In this way one obtains the NNLO
expression for $\delta_{\tau}$ which has the form (\ref{R}) with the
expansion coefficients $\tilde{r}_{i}$:
\begin{equation}
\tilde{r}_{1}=r_{1}+\frac{19}{24}\,b,
\end{equation}
\begin{equation}
\tilde{r}_{2}=r_{2}+\frac{19}{12}\,b\,r_{1}+\frac{19}{24}\,b\,c_{1}+
\frac{265-24\pi^{2}}{288}\,b^{2}.
\end{equation}
In the $\overline{MS}$ scheme we have \cite{kataev}:
\begin{equation}
\tilde{r}_{1}^{\overline{MS}}=5.2023, \,\,\,\,
\tilde{r}_{2}^{\overline{MS}}=26.366.
\end{equation}
We see that there is a significant difference between the expansion
coefficients $r_{i}$ and $\tilde{r}_{i}$.

It should be noted however, that using the expansion of $a(-s)$ in
terms of $a(m_{\tau}^{2})$ in the NNLO expression for $\delta_{\Pi}$,
 we make a
rather crude approximation. For example, we effectively assume, that:
\begin{equation}
a^{3}(-s)\approx a^{3}(m^{2}_{\tau}).
\end{equation}
 As was pointed out in \cite{piv,ledepi} one may obtain a more
accurate description of the QCD effects in the tau lepton decay by
evaluating the $s$-dependence of the running coupling constant on the
contour $C$ in a more precise way.

It is then an interesting question, whether the improved evaluation of
the QCD corrections may change the rather upsetting result found in
\cite{b}.

In this note we analyze the predictions for $\delta_{\tau}$ obtained from
the contour integral representation (\ref{rtcont}) using a numerical
solution of the renormalization group equation in the complex energy
plane. To obtain the running coupling constant along the contour $C$
we solve numerically the implicit equation:
\begin{equation}
\frac{b}{2}\ln\left(\frac{m^{2}_{\tau}}
{\Lambda^{2}_{\overline{MS}}}\right)+
i\frac{b\theta}{2}=r^{\overline{MS}}_{1}-r_{1}+\Phi(a,c_{2}),
\end{equation}
where
\begin{equation}
\Phi(a,c_{2})=c_{1}\ln\left(\frac{b}{2c_{1}}\right)+
\frac{1}{a}+c_{1}\ln(c_{1}a)+O(a).
\end{equation}
The explicit form of $\Phi(a,c_{2})$ is given for example in
\cite{rge}. (The renormalization group coefficients for $n_{f}=3$ are
$b=4.5$, $c_{1}=16/9$ and
$c_{2}^{\overline{MS}}=3863/864\approx4.471$.)  This equation is
obtained by integrating the renormalization group equation (\ref{rge})
with an appropriate boundary condition \cite{msb}, and then
analytically continuing to general complex $s$. (The presence of
$r_{1}^{\overline{MS}}$ and $\Lambda_{\overline{MS}}$
in this general expression results from taking
explicitly into account the one-loop relation between $\Lambda$
parameters in different schemes \cite{CelGon}, which is valid to all
orders of perturbation expansion. All values of $\Lambda_{\overline{MS}}$
mentioned in this paper correspond to three ``active'' quark flavors.)
  To obtain the prediction for
$\delta_{\tau}$ we take under the contour integral the quantity $\delta_{\Pi}$
in the NNLO approximation in the form (\ref{R}) and we perform the
contour integral numerically.   To analyze the RS dependence of the
thus obtained expression for $\delta_{\tau}$ we parametrize the freedom of
choice of the RS by the coefficient $r_{1}$ in $\delta_{\Pi}$ and the
coefficient $c_{2}$, and we calculate variation in the predictions
when these parameters are changed in a region determined by the
condition (\ref{constraint}).  The region of allowed values for
$r_{1}$ and $c_{2}$ for $\rho_{2}>c_{1}^{2}/4$, which is the case
needed here, was described analytically in \cite{c}.

 In our analysis a fundamental role is played by the NNLO RS
invariant, which for $\delta_{\Pi}$ has the value $\rho_{2}=5.238$. This
value should be compared with $\tilde{\rho_{2}}=-5.476$, which is
obtained for the conventional expansion of $\delta_{\tau}$ in terms of
$a(m^{2}_{\tau})$.  The fact that $\tilde{\rho}_{2}$ has similar
magnitude, but the opposite sign,  compared to $\rho_{2}$ indicates that
the approximations used to derive the conventional expansion for
$\delta_{\tau}$ may greatly distort the pattern of the RS dependence of the
predictions.

 The results of our calculations are presented in four figures.
In Fig.1 we compare perturbative QCD predictions for $\delta_{\tau}$
 as a function of $m_{\tau}/\Lambda_{\overline{MS}}$, obtained
 in the next-to-leading order (NLO) and in
 NNLO using the conventional expansion, with the predictions obtained
 by evaluating numerically the contour integral.
 We clearly see
 that the differences between NLO and NNLO are  much smaller
 for the predictions obtained from numerical calculation of the
 contour integral than for the predictions obtained with the
 conventional expansion in terms of $a(m_{\tau}^{2})$. This
 confirms the expectation that the procedure for evaluating
 the $\delta_{\tau}$ adopted in this note gives indeed improvement
 over the conventional expansion.

  In Fig.2 we show a contour plot for
 $\delta_{\tau}$ as a function of $r_{1}$
and $c_{2}$, obtained for $\Lambda_{\overline{MS}}=310\,$~MeV,
together with the region of scheme parameters satisfying the condition
(\ref{constraint}) with $l=2$ --- smaller region --- and $l=3$.  (Note
that --- for purely technical reasons --- the actual variable on the
vertical axis is $c_{2}-c_{1}r_{1}$ instead of $c_{2}$.)  We see that
the pattern of RS dependence in this case has more complicated
structure than that obtained for conventional NNLO approximants ---
this is evident if one compares Fig.2 with the corresponding figure in
\cite{c}, representing the RS dependence of the QCD corrections to the
Bjorken sum rule. In particular, we find here {\em three} critical
points which are relatively close to the allowed region, instead of
one encountered in the conventional approximant.   The coordinates of
the critical point closest to the origin, i.e.\ the saddle point at
the boundary of the $l=2$ allowed region, are very close to the PMS
parameters for the quantity $\delta_{\Pi}$ evaluated for real negative~$s$
 (``euclidean'' PMS). (Let us note that
because of a nonpolynomial character of the improved expression for
$\delta_{\tau}$
the value of the PMS parameters cannot be obtained from the set of
 algebraic equations given in \cite{PMS}, but it has to be
 determined from direct numerical analysis.)
 This confirms the fundamental importance of $\delta_{\Pi}$ for
$\delta_{\tau}$.
It also shows that  approximations used to derive the conventional
NNLO expression greatly distort the pattern of the  RS dependence --- the
conventional NNLO approximant for $\delta_{\tau}$ would have a critical
point for completely different values of the scheme parameters.
To estimate  the strength of the RS dependence we compare
predictions for $\delta_{\tau}$ obtained for the scheme
parameters lying in the denoted allowed region.
 Because of the location of
the critical points in this case the maximal and minimal values for
$\delta_{\tau}$ are attained at the boundaries of the allowed regions
corresponding to $l=2$ and $l=3$. It is interesting to note that the
$\overline{MS}$ scheme with $c_{2}-c_{1}r_{1}\approx1.56$
 lies slightly outside the $l=3$ allowed region.

 In Fig.3 we show, how the pattern of the RS dependence of the
improved expression for $\delta_{\tau}$ depends on the value of
$\ln(m_{\tau}/\Lambda_{\overline{MS}})$.
 We see that for small values of
$\Lambda_{\overline{MS}}$ we have only one critical point in the
 $l=2$ allowed region, and that it lies  very close to the
 PMS parameters for
$\delta_{\Pi}$ evaluated for spacelike momenta. With increasing
$\Lambda_{\overline{MS}}$ the
 structure of the critical points close
to the $l=2$ allowed region becomes more complicated.
 For some values of $\Lambda_{\overline{MS}}$ three
critical points appear. For large values of
$\Lambda_{\overline{MS}}$ the critical point corresponding to the
``euclidean'' PMS evolves into a very flat plateau,
and the secondary critical point,
with negative $r_{1}$, moves  into the $l=2$ allowed region.

 In Fig.4, which is the main result of this paper, we show how the
optimal perturbative predictions for $\delta_{\tau}$,
 and the estimates of the RS
dependence, behave as a function of
$m_{\tau}/\Lambda_{\overline{MS}}$. The results of our
calculations are compared with  constraints from the  experimental data.
As our preferred predition, indicated in Fig.4 by a thick solid line,
we take the values obtained in the scheme with
$r_{1}=0$, $c_{2}=1.5\rho_{2}$. These parameters describe,
for small and intermediate
 values of $\Lambda_{\overline{MS}}$, the approximate location of the
 leading critical point in the $(r_{1},c_{2})$ plane, and for larger
values of $\Lambda_{\overline{MS}}$ they fall in the region of
very small variation of the predictions.
{}From Fig.3 it is clear, that for all interesting  values of
 $\Lambda_{\overline{MS}}$ the  predictions for $\delta_{\tau}$
 in this scheme practically coincide with the
  values at the relevant  critical points.
 The dashed and dash-dotted lines represent the variation of the
 predictions over the $l=2$ and $l=3$ allowed regions, respectively.
In order to make our results useful for other authors we give
 in Table~1 the
 numerical values of the preferred predictions for $\delta_{\tau}$,
 together with an  estimate of the RS dependence,
 for several values of $\ln(m_{\tau}/\Lambda_{\overline{MS}})$.
For completeness we include also the NLO
predictions, evaluated using the contour integral and
 optimized according the the PMS prescription --- they
are well approximated by taking the scheme with $r_{1}=-0.76$.

  We see that the variation of the predictions over the
$l=2$ allowed region has a rather smooth  dependence on
$\Lambda_{\overline{MS}}$, being almost
constant for the considered range of $\Lambda_{\overline{MS}}$. The
variation over the $l=3$ region of parameters is reasonably close to
the $l=2$ variation, except for the maximal value, which grows rapidly
for larger  $\Lambda_{\overline{MS}}$.  This should be compared with the
large differences found in \cite{b} for the conventional NNLO
approximant for $\delta_{\tau}$. We see that using an improved evaluation
of the predictions for $\delta_{\tau}$ we  qualitatively
improve stability of the predictions with respect to
 change of the RS, and we greatly reduce the ambiguities resulting from the
freedom of choice of the renormalization scheme.

 Let us now compare our results with those obtained in  \cite{ledepi}.
 The  analysis of \cite{ledepi} concentrates on variation of the
 renormalization scale $\mu$ in the $\overline {MS}$ scheme, which
 in our approach corresponds to variation of $r_{1}$ for fixed
 $c_{2}=c_{2}^{\overline{MS}}$. In \cite{ledepi}
 only a brief comment is made on the variation of predictions
 when $c_{2}$ is changed from $0$ to $2c_{2}^{\overline{MS}}$
 with fixed $\mu$.
 It must be  stressed, that  there is
 no theoretical or phenomenological reason justifying
 {\it a priori}
 such a  choice of the set of the scheme parameters.
 Also,  there is no immediate relation between the magnitude of
  variation found according to \cite{ledepi} and the estimate of
 the RS dependence obtained according to our method. In particular,
 small RS dependence obtained
 by prescription of \cite{ledepi} does not necessarily imply
 a small variation in our approach.  It should be emphasized, that
 if we agree that there is a ``democracy'' of renormalization
 schemes, then the set of considered  schemes  must be  larger than
 that discussed in \cite{ledepi}, and
 any reasonable constraint on the freedom of choice
 must be related to the RS invariant $\rho_{2}$.
 Most importantly,
 a consideration of an essentially
 two-dimensional set of scheme parameters is unavoidable if one
 intends to find the optimal NNLO predictions according to the
 Principle of Minimal Sensitivity, which
 requires for the optimal scheme the vanishing of the
 derivatives with respect to {\em both}
 scheme parameters.

 The results  presented  in Fig.4 may be used  to obtain
 an improved constraint on  $\Lambda^{(3)}_{\overline{MS}}$ from the
 world average of the  available experimental data on $R_{\tau}$.
 As is well known, the experimental value
 of $R_{\tau}$ may be obtained in three independent  ways: from the
 electronic and muonic branching ratios in the tau lepton
 decay, and from the total width of the tau lepton.  (Note
 that Particle Data Group gives {\em two} sets of
  values for the leptonic  branching fractions --- the number
 shown below corresponds to the set called ``our average.''
 See Appendix for a detailed discussion of the subtleties
 involved in obtaining experimental value for $R_{\tau}$.)
 Taking a weighted average of these three determinations we find:
\begin{equation}
(R_{\tau})_{exp}^{avg}=3.591\pm0.036.
\label{Rtex}
\end{equation}

 This value may then be converted into experimental constraint
 on $\delta_{\tau}$ via Eq.(\ref{rt}). However, in order to have
 a phenomenologically meaningful constraint on $\delta_{\tau}^{pert}$, we
 must take into account also the effect of the light quark masses
 and the nonperturbative effects. According to
 \cite{branapi,chetkwiat} these may be estimated
 to give roughly  an overall $-(1.5\pm0.4)\%$ correction to
 the value of $R_{\tau}$. Therefore, after taking into
 account  that the factor involving the  CKM matrix elements
 is practically equal to unity,
  our final formula relating
 $R_{\tau}$ and $\delta_{\tau}^{pert}$ takes
 the form:
\begin{equation}
R_{\tau}=3\times1.0194
\times(1.001+\delta_{\tau}^{pert})\times(0.985\pm0.004).
\end{equation}
 Using (\ref{Rtex}) we obtain:
\begin{equation}
(\delta_{\tau}^{pert})_{exp}=0.191\pm0.012(\mbox{exp})
\pm0.005(\mbox{npt}),
\end{equation}
where the first error reflects the effect of experimental
 uncertainties and  the second one the effect of uncertainties in
  nonperturbative corrections.

Fitting $\Lambda_{\overline{MS}}^{(3)}$ to this experimental
 value we obtain:
\begin{equation}
\Lambda^{(3)}_{\overline{MS}}=376(\mbox{opt})
^{+15}_{-14}(\mbox{th,l=2})\pm29(\mbox{exp})
\pm12(\mbox{npt})\,\mbox{MeV}.
\end{equation}
For the $l=3$ region of parameters the variation in
$\Lambda_{\overline{MS}}^{(3)}$ is $^{+26}_{-21}\,\mbox{MeV}$.

As was explained in \cite{b,c}, a natural way to parametrize the
 QCD predictions in our analysis is via the $\Lambda_{\overline{MS}}$.
 However, for  comparison with results given in other papers on
  experimental tests of QCD,  we give also the
corresponding values of $\alpha_{s}(m_{\tau}^{2})$ in the
$\overline{MS}$ scheme, obtained with the three-loop renormalization
group equation:
\begin{equation}
\alpha_{s}^{\overline{MS}}(m_{\tau}^{2})=
0.332^{+0.008}_{-0.007}(\mbox{th,l=2})
\pm0.015(\mbox{exp})\pm0.006(\mbox{npt}).
\end{equation}
For the $l=3$ region of parameters we obtain for
$\alpha_{s}^{\overline{MS}}(m_{\tau}^{2})$ the variation of
$^{+0.014}_{-0.010}$.

Using the formulas given in  \cite{marciano} to match
$\Lambda_{\overline{MS}}$ for different number of flavors
we obtain the corresponding
result for  $\alpha_{s}(m_{Z}^{2})$:
\begin{equation}
\alpha_{s}^{\overline{MS}}(m_{Z}^{2})=
0.1190^{+0.0009}_{-0.0008}(\mbox{th,l=2})
\pm0.0017(\mbox{exp})\pm0.0007(\mbox{npt}).
\end{equation}
For the $l=3$ region of parameters we obtain for
$\alpha_{s}^{\overline{MS}}(m_{Z}^{2})$ the variation of
$^{+0.0016}_{-0.0013}$.

We see that the RS
dependence ambiguities in the determination of
$\alpha_{s}^{\overline{MS}}(m_{\tau}^{2})$
 or $\alpha_{s}^{\overline{MS}}(m_{Z}^{2})$
 from $R_{\tau}$ are
 not very big --- this is the result of stabilization of
the predictions when improved evaluation procedure is used.
 Note however that the RS dependence ambiguities are still
 comparable in magnitude to the
 uncertainties related to  the
 present experimental accuracy of $R_{\tau}$.

Let us note that using for the leptonic branching fractions the
values denoted by PDG  as ``our fit'' \cite{pdg} we obtain
(see Appendix):
\begin{equation}
(R_{\tau})_{exp}^{avg}=3.535\pm0.037.
\end{equation}
which is $1.5\sigma$ smaller than (\ref{Rtex}). Using this value
we find in NNLO:
\begin{equation}
\Lambda^{(3)}_{\overline{MS}}=331\mbox{MeV}(\mbox{opt}),\,\,\,\,\,
 \alpha_{s}^{\overline{MS}}(m_{\tau}^{2})=0.309,\,\,\,\,\,
\alpha_{s}^{\overline{MS}}(m_{Z}^{2})=0.1162,
\end{equation}
with the same theoretical and experimental uncertainties as before.

To have a complete picture of theoretical uncertainties we give
also the result of a fit of $\Lambda_{\overline{MS}}$ to the experimental
value of $R_{\tau}$, obtained with the improved NLO
predictions optimized according to the PMS prescription.
 Using the NLO expression with $r_{1}=-0.76$ we obtain:
\begin{equation}
\Lambda^{(3)}_{\overline{MS}}=413(\mbox{opt})
\pm32(\mbox{exp})\,\mbox{MeV}.
\end{equation}
Using then the two-loop renormalization group equation we find:
\begin{equation}
\alpha_{s}^{\overline{MS}}(m_{\tau}^{2})=
0.316\pm0.013(\mbox{exp}),
\end{equation}
\begin{equation}
\alpha_{s}^{\overline{MS}}(m_{Z}^{2})=
0.1209\pm0.0018(\mbox{exp}).
\end{equation}

Summarizing, we have evaluated the perturbative QCD corrections to the
semileptonic decay width of the tau lepton,
 using the contour integral
representation and evaluating numerically the running coupling constant
 on the contour. We obtained results in a broad class of
renormalization schemes.
We discussed the problem of obtaining the optimal predictions
in the improved evaluation, and  we found  predictions
for $\delta_{\tau}$ in
the scheme preferred by the Principle of Minimal Sensitivity, both
in NLO and in NNLO.
 Using a specific condition to eliminate the schemes
with unnaturally large expansion coefficients we obtained a
quantitative estimate of the ambiguities in $\delta_{\tau}$ arising from
the freedom of choice of the RS. We expect that this estimate would
 be useful in combining the QCD results from the $R_{\tau}$ with
 other experimental constraints on QCD.
  We found that the improved expression
for $\delta_{\tau}$ is much more stable with respect to
 change of RS compared to
the conventional perturbative NNLO expression. Nevertheless, we found
that the RS dependence ambiguities in the fit of
$\Lambda^{(3)}_{\overline{MS}}$ to the experimental data for
$\delta_{\tau}$ are at least of the order of uncertainties arising
from the present  accuracy of  experimental determinations
of $R_{\tau}$. \vspace{1cm}

\noindent {\it Note added}\\
\nopagebreak
Soon after this paper was posted on the bulletin board, two
papers appeared \cite{pich94} \cite{narison94} in which determination
of $\alpha_{s}$ from the tau decay is discussed in detail.
We make two comments about these papers. First,
the value of $\alpha_{s}$ obtained in \cite{pich94} is almost identical to
the value given in our paper. However, this value was
obtained using $(R_{\tau})_{exp}=3.56\pm0.03$,  which is $1\sigma$
lower than the value we have used in our analysis. Secondly,
the main source of the theoretical uncertaintly estimated
in \cite{pich94} \cite{narison94}
appears to be the contribution from the yet uncalculated
$O(\alpha_{s}^{4})$ corrections. Such an estimate necessarily
involves some speculative assumptions on the magnitude of the
higher order terms.  Also, the higher order contributions would
be RS dependent, with the freedom of choice of the RS characterized
by {\em three} arbitrary parameters in the four-loop order.
It is straightforward to
extend the method used in our paper to the four-loop  case
in order
to obtain a quantitative estimate of the resulting RS ambiguity.

\section*{Appendix}

In this appendix we make some remarks on the experimental
determination of  $R_{\tau}$. First, let us note that
the experimental value
 of $R_{\tau}$ may be obtained  from {\em three}
independent measurements: from the
 electronic and muonic branching ratios in the tau lepton
 decay, and from the total width of the tau lepton. To
 obtain $R_{\tau}$ from the electronic branching ratio we
 take:
\begin{displaymath}
(R_{\tau})_{exp}^{B_{e}}=\frac{1}{(B_{e})_{exp}}-1-
\left(\frac{B_{\mu}}{B_{e}}\right)_{th},
\end{displaymath}
where  according to formulas given in \cite{marsir}, with
 slightly updated parameters, we should take:
\begin{displaymath}
\left(\frac{B_{\mu}}{B_{e}}\right)_{th}=0.972568.
\end{displaymath}
 To
 obtain $R_{\tau}$ from the muonic branching ratio we
 take:
\begin{displaymath}
(R_{\tau})_{exp}^{B_{\mu}}=\frac{1}{(B_{\mu})_{exp}}
\left(\frac{B_{\mu}}{B_{e}}\right)_{th}
-1-\left(\frac{B_{\mu}}{B_{e}}\right)_{th}.
\end{displaymath}
Finally,  to
 obtain $R_{\tau}$ from the total decay width we
 take:
\begin{displaymath}
(R_{\tau})_{exp}^{\Gamma_{\tau}}=
\frac{(\Gamma_{tot})_{exp}}{(\Gamma_{e})_{th}}-1-
\left(\frac{\Gamma_{\mu}}{\Gamma_{e}}\right)_{th},
\end{displaymath}
with
\begin{displaymath}
(\Gamma_{e})_{th}=(0.40341\pm0.00057)\times10^{-12}\mbox{GeV},
\end{displaymath}
obtained from the formulas given in \cite{marsir} with updated
 parameters.

As may be expected, these three determinations do not give
exactly the same value. Therefore we take for $R=<R>\pm\delta R$
 a weighted average, according to the prescription:
\begin{displaymath}
<R>=\frac{\sum_{i}w_{i}R_{i}}{\sum_{i}w_{i}},\,\,\,
\delta R=(\sum_{i}w_{i})^{-1/2},
\end{displaymath}
where $w_{i}=(\delta R_{i})^{-2}$.

Secondly, one has to be careful which set of the  world averaged
experimental data one uses in the phenomenological analysis.
The Particle Data Group \cite{pdg} gives in fact {\em two}
sets of values for the leptonic branching fractions.
One set  is a straightforward
world average of the data:
\begin{displaymath}
(B_{e})_{exp}=0.1790\pm0.0017,\,\,\,\,\,
(B_{\mu})_{exp}=0.1744\pm0.0023,
\end{displaymath}
This set is  called by PDG ``our average.'' Another set,
\begin{displaymath}
(B_{e})_{exp}=0.1801\pm0.0018,\,\,\,\,\,
(B_{\mu})_{exp}=0.1765\pm0.0024,
\end{displaymath}
called ``our fit,''
is a result of a global fit to the tau lepton branching
fractions, taking into account the relevant constraints.
Taking ``our average'' of PDG for the leptonic branching fractions
we obtain:
\begin{displaymath}
(R_{\tau})_{exp}^{B_{e}}=3.614\pm0.053,\,\,\,\,\,
(R_{\tau})_{exp}^{B_{\mu}}=3.604\pm0.074.
\end{displaymath}
Taking the present world average \cite{pdg} for the total
decay width:
\begin{displaymath}
(\Gamma_{tot})_{exp}=\frac{\hbar}{(\tau_{\tau})_{exp}}=
(2.227\pm0.023)\times10^{-12}\mbox{GeV},
\end{displaymath}
we find:
\begin{displaymath}
(R_{\tau})_{exp}^{\Gamma_{\tau}}=3.548\pm0.065.
\end{displaymath}
(For $\tau_{\tau}$ the PDG also gives two values ---
in this case it seems however clear that one should take the value
in which results of older experiments have been corrected for the
change in the experimental number for $m_{\tau}$.)
Taking a weighted average of three values for $R_{\tau}$ we find:
\begin{displaymath}
(R_{\tau})_{exp}^{avg}=3.591\pm0.036.
\label{average}
\end{displaymath}
This value has been further used in our paper to obtain constraints
on $\Lambda_{\overline{MS}}^{(3)}$.

It has to be emphasized however, that if we use ``our fit''
of PDG for the leptonic branching fractions,  we find:
\begin{displaymath}
(R_{\tau})_{exp}^{B_{e}}=3.525\pm0.055,\,\,\,\,\,
(R_{\tau})_{exp}^{B_{\mu}}=3.538\pm0.075.
\end{displaymath}
Together with the value obtained from the total decay width this
gives a weighted average of:
\begin{displaymath}
(R_{\tau})_{exp}^{avg}=3.535\pm0.037.
\end{displaymath}

\newpage
\section*{Figure Captions}

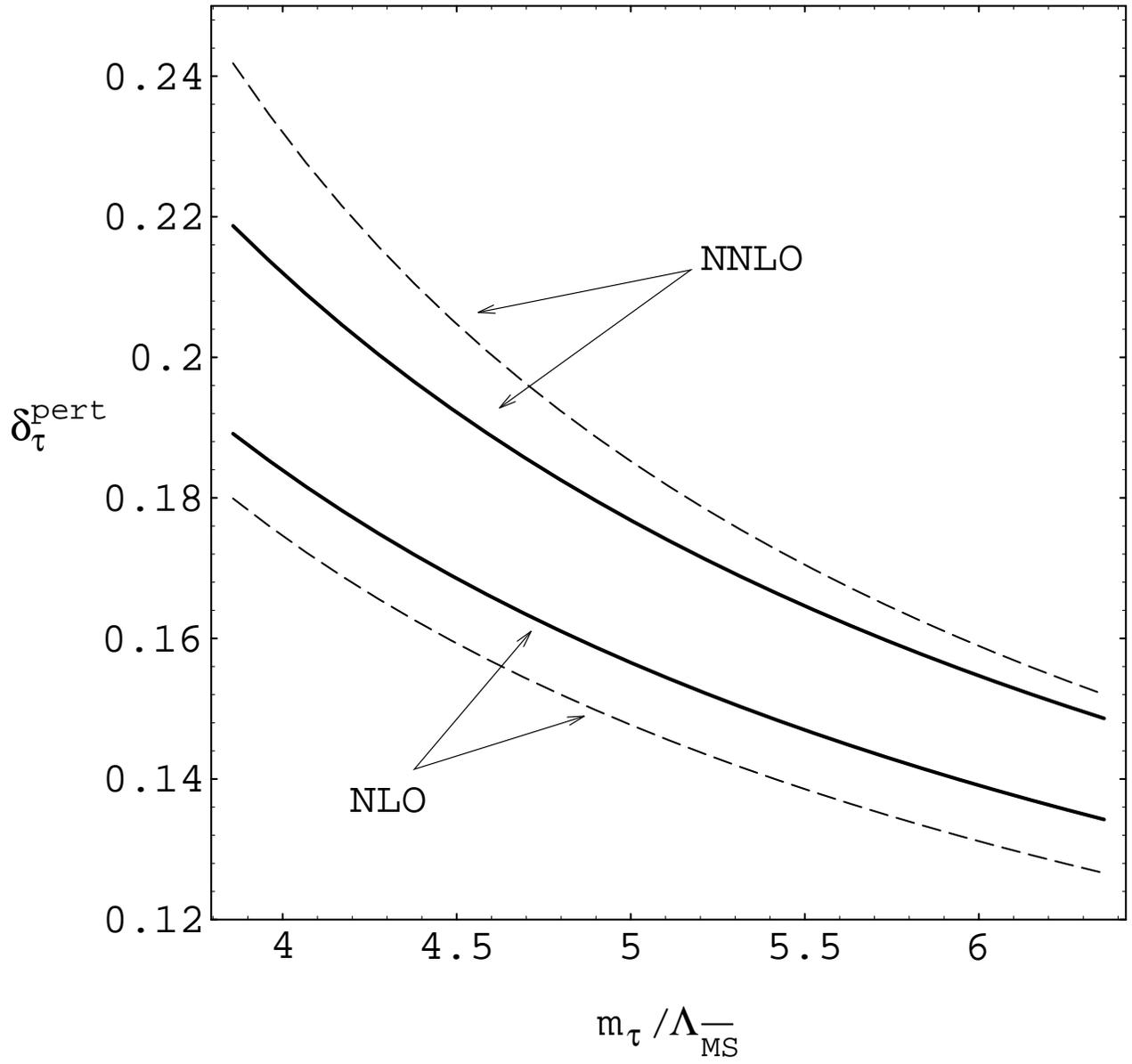
\noindent Fig.1. Perturbative QCD predictions for $\delta_{\tau}$
 as a function of $m_{\tau}/\Lambda^{(3)}_{\overline{MS}}$, obtained
 in NLO and NNLO with the  conventional expansion (dashed lines)
 and with the  numerical evaluation of the contour integral
 (solid lines).
\vspace{.5cm}

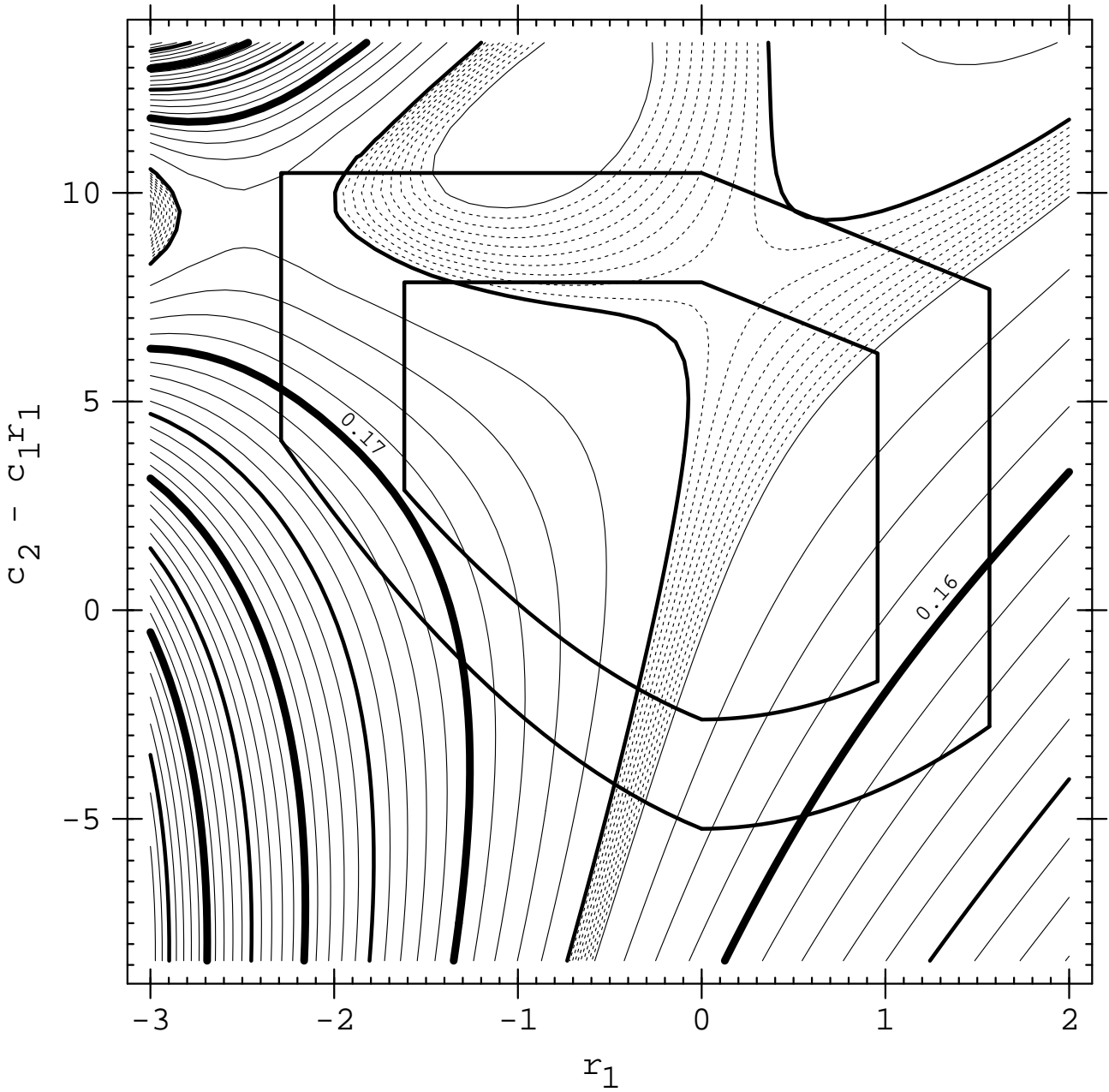
\noindent Fig.2. Perturbative prediction for $\delta_{\tau}$, obtained from
 the  numerical evaluation of the contour integral
 for $\Lambda^{(3)}_{\overline{MS}}=310\,$~MeV,
 as a function of the parameters
 $r_{1}$ and $c_{2}$ determining the RS.
 Note that the actual variable on the vertical axis is $c_{2}-c_{1}r_{1}$.
 The regions
 of scheme parameters satisfying the condition (\ref{constraint})
 with $l=2$ and $l=3$ are also indicated. At the critical point inside
the $l=2$ region we have $\delta_{\tau}=0.1649$.
\vspace{.5cm}

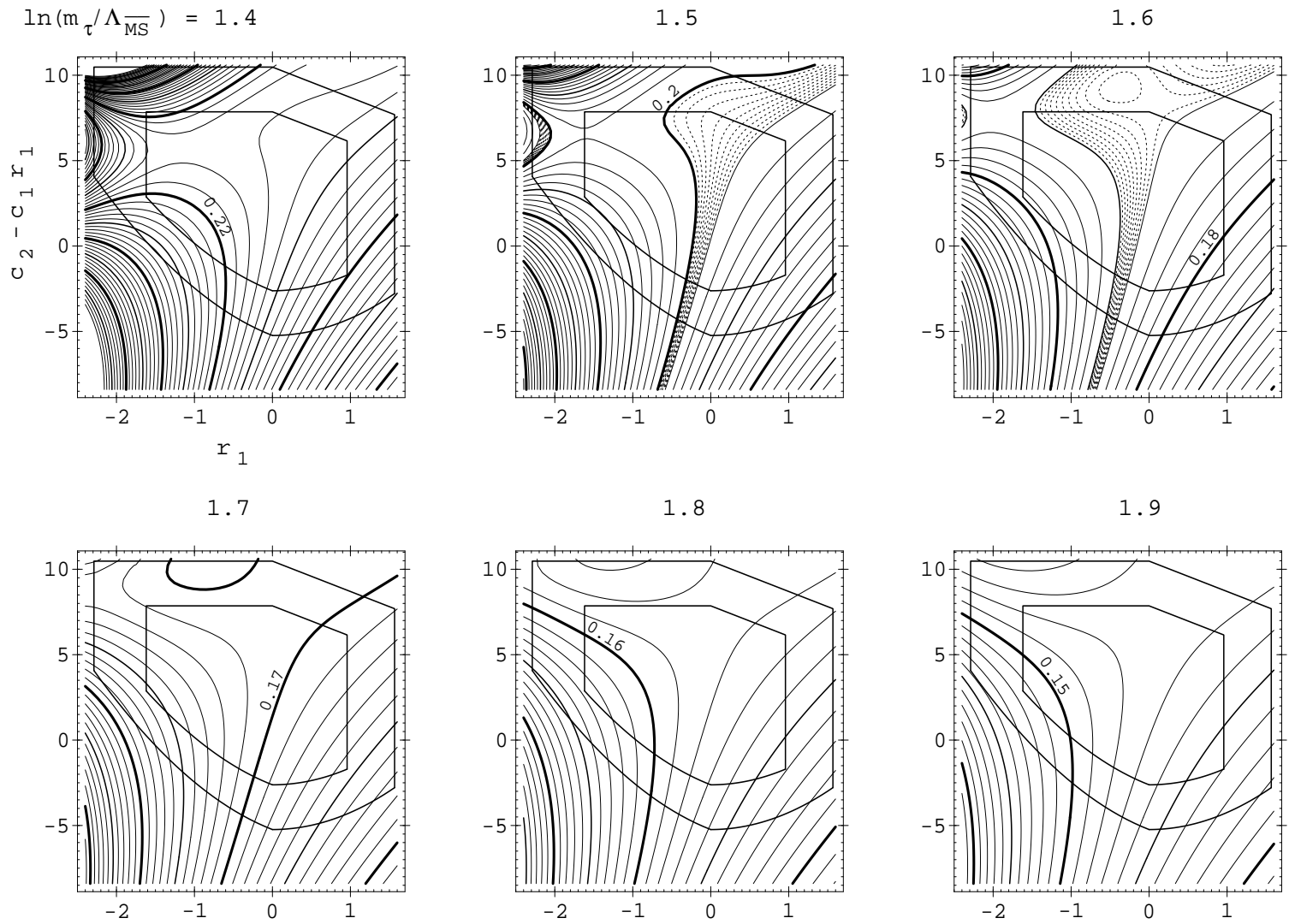
\noindent Fig.3. Same as in Fig.2, but for various values of
 $\ln(m_{\tau}/\Lambda^{(3)}_{\overline{MS}})$.
\vspace{.5cm}

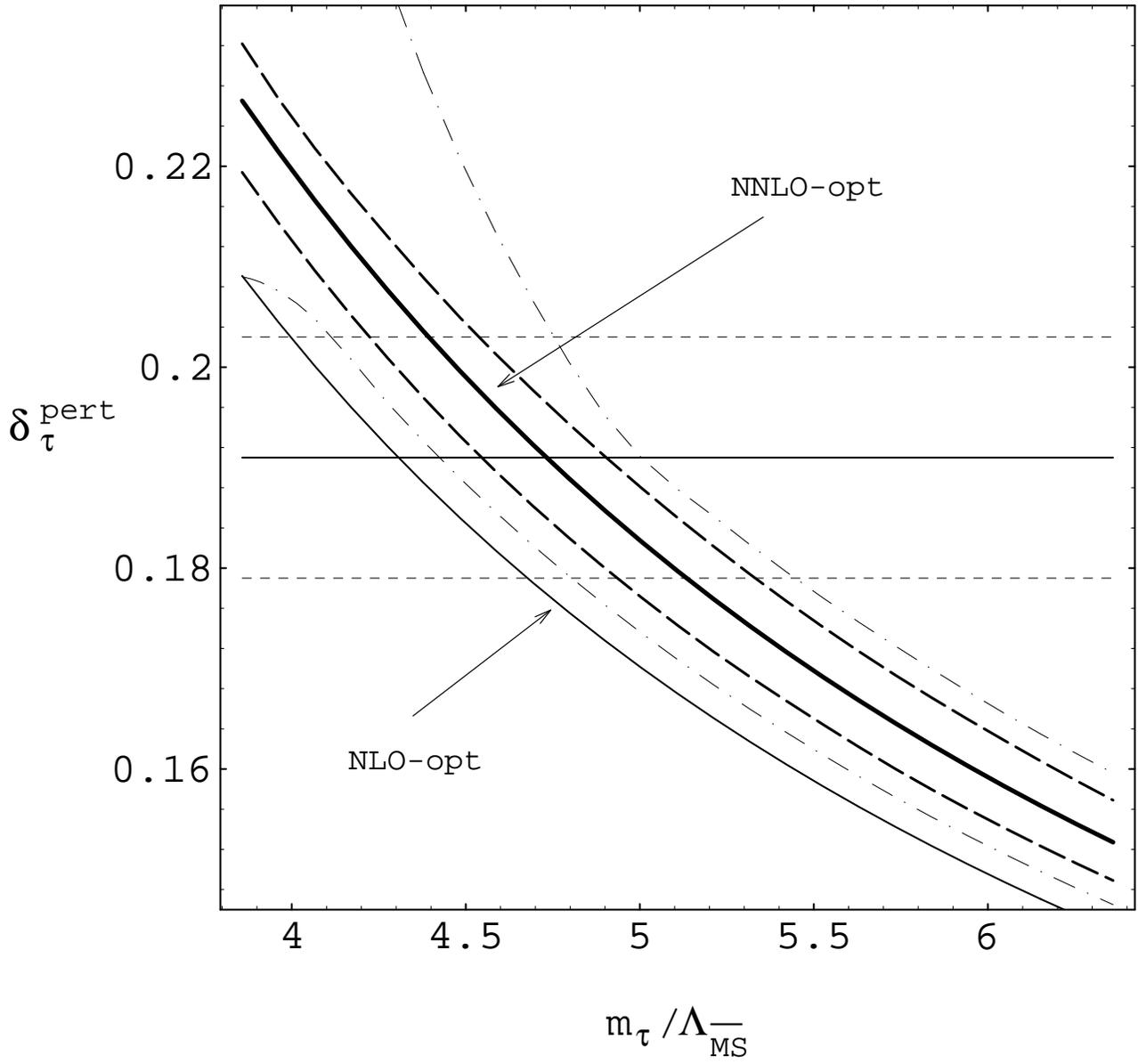
\noindent Fig.4. Perturbative prediction for $\delta_{\tau}$, obtained
 from the  numerical evaluation of the contour integral
 representation, as a function of
 $m_{\tau}/\Lambda^{(3)}_{\overline{MS}}$.
 The thick lines denote the preferred NNLO predictions
(i.e. obtained in  the scheme with $r_{1}=0$ and
 $c_{2}=1.5\rho_{2}$ --- see text)
 and the optimal NLO predictions (obtained
 with $r_{1}=-0.76$).
 Also indicated are the maximal and minimal values
 obtained in NNLO when the scheme parameters are
 varied over the $l=2$ (dashed lines)
 and $l=3$ (dash-dotted lines) allowed regions. For
 comparison we show the experimental constraint on the
 value of $\delta_{\tau}^{pert}$.

\newpage
\begin{table}
\begin{center}
\begin{tabular}{||c|l|l|l|l|l|l||}
\hline
$\ln(m_{\tau}/\Lambda^{(3)}_{\overline{MS}})$
& $(\delta_{\tau})_{NNLO}^{opt}$ &
$\delta^{min}_{l=2}$ & $\delta^{max}_{l=2}$ &
$\delta^{min}_{l=3}$ & $\delta^{max}_{l=3}$&
 $(\delta_{\tau})_{NLO}^{opt}$  \\
\hline
1.30 & .23634 & .2293 & .2443 & .2075 & --- & .21816  \\
\hline
1.35 & .22652 & .2194 & .2322 & .2090 & --- & .20909 \\
\hline
1.40 & .21711 & .2101 & .2223 & .2052 & --- & .20056 \\
\hline
1.45 & .20816 & .2014 & .2135 & .1969 & .2401 & .19256\\
\hline
1.50 & .19968 & .1932 & .2050 & .1890 & .2209 & .18505\\
\hline
1.55 & .19168 & .1856 & .1971 & .1817 & .2055 & .17800\\
\hline
1.60 & .18414 & .1785 & .1895 & .1749 & .1928 & .17139\\
\hline
1.65 & .17705 & .1718 & .1823 & .1685 & .1853 & .16518\\
\hline
1.70 & .17039 & .1656 & .1754 & .1625 & .1783 & .15935\\
\hline
1.75 & .16413 & .1596 & .1689 & .1568 & .1718 & .15387\\
\hline
1.80 & .15824 & .1541 & .1628 & .1515 & .1655 & .14871\\
\hline
1.85 & .15271 & .1489 & .1569 & .1465 & .1596 & .14386\\
\hline
1.90 & .14751 & .1440 & .1514 & .1418 & .1540 & .13928\\
\hline
\end{tabular}
\end{center}
\caption{
 Numerical values of the preferred NNLO  predictions
 for $\delta_{\tau}$ (i.e. obtained in the  scheme with
  $r_{1}=0$ and $c_{2}=1.5\rho_{2}$
 --- see text),  and the maximal and minimal values
 obtained after variation of the scheme parameters
 in the  allowed region corresponding to  $l=2$ and
 $l=3$, for
 several values of $\ln(m_{\tau}/\Lambda^{(3)}_{\overline{MS}})$.
 For comparison also the values of the preferred
 NLO predictions are given ($r_{1}=-0.76$).
 }
\end{table}


\begin{thebibliography}{99}

\bibitem{b} P. A. R\c{a}czka,  Phys. Rev. {\bf D 46}
 R3699  (1992), see also P. A. R\c{a}czka, Proceedings of the XV
International Warsaw Meeting on Elementary Particle Physics,
Kazimierz, Poland, 25-29 May 1992, edited by Z. Ajduk, S. Pokorski and
A. K. Wr\'{o}blewski (World Scientific, Singapore, 1993), p.\ 496.

\bibitem{c} P. A. R\c{a}czka, Zeitschrift f\"ur Physik {\bf C65},
 481 (1995).

\bibitem{lamyan} C. S. Lam and T. M. Yan, Phys. Rev. {\bf D16},
 703 (1977).

\bibitem{schil} K. Schilcher and M. D. Tran, Phys. Rev.
 {\bf D29}, 570 (1984).

\bibitem{bra88} E. Braaten, Phys. Rev. Lett. {\bf 60}, 1606 (1988),
 ibid. {\bf 63}, 577 (1989).

\bibitem{narpi} S. Narison and A. Pich, Phys. Lett. {\bf B211}, 183 (1988).

\bibitem{bra89} E. Braaten, Phys. Rev. {\bf D39}, 1458 (1989).

\bibitem{kataev} S. G. Gorishny, A. L. Kataev and S. A. Larin,
 Phys. Lett. {\bf B259}, 144 (1991).

\bibitem{samuel} M. A. Samuel and L. R. Surguladze, Phys. Rev.
 {\bf D44}, 1602 (1991).

\bibitem{branapi} E. Braaten, S. Narison and A. Pich, Nucl. Phys.
 {\bf B373}, 581 (1992).

\bibitem{piv} A. A. Pivovarov, Z. Phys. {\bf C53}, 461 (1992).

\bibitem{ledepi} F. LeDiberder and A. Pich, Phys. Lett. {\bf B286}, 147
 (1992).

\bibitem{rszym} Results reported here have been presented at the
 XVII Kazimierz Meeting on
 Elementary Particle Physics, Kazimierz, Poland, May 23-27, 1994.

\bibitem{msb} W. A. Bardeen, A. J. Buras, D. W. Duke, and
 T. Muta, Phys. Rev. {\bf D18}, 3998  (1978).

\bibitem{CelGon} W. Celmaster and R. J. Gonsalves,
 Phys. Rev. {\bf D20}, 1426  (1979).

\bibitem{GrunFAC} G. Grunberg, Phys. Lett. {\bf 95B}, 70  (1980).

\bibitem{PMS} P. M. Stevenson,  Phys. Lett. {\bf 100B}, 61 (1981),
 Phys. Rev. {\bf D23}, 2916  (1981).

\bibitem{CelSiv} W. Celmaster and D.  Sivers, Phys.\ Rev.\
 {\bf D23}, 227 (1981).

\bibitem{DharGuptaMOM} A. Dhar and V. Gupta, Phys. Lett. {\bf 101B},
 432 (1981).

\bibitem{BLM} S. J. Brodsky, G. P. Lepage,  and P. Mackenzie,
 Phys. Rev. {\bf D28}, 228 (1983).

\bibitem{Dhar} A. Dhar, Phys. Lett. {\bf 128B}, 407 (1983).

\bibitem{DharGupta} A. Dhar and V. Gupta,  Pram\={a}na {\bf 21},
 207 (1983), Phys. Rev. {\bf D29}, 2822 (1984).

\bibitem{GrunEC} G. Grunberg,  Phys. Rev. {\bf D29}, 2315 (1984).

\bibitem{DukeRoberts} For a summary of early contributions see
 D. W. Duke and R. G. Roberts, Phys. Rep. {\bf 120}, 275 (1985).

\bibitem{marsir} W. J. Marciano and A. Sirlin, Phys. Rev. Lett.
 {\bf 61}, 1815 (1988).

\bibitem{brali} E. Braaten and Chong Sheng Li, Phys. Rev.
 {\bf D42}, 3888 (1990).

\bibitem{maxnich} C. J. Maxwell and J. A. Nicholls, Phys. Lett.
 {\bf B236}, 63 (1990).

\bibitem{haru90} M. Haruyama, Prog. Theor. Phys. {\bf 83}, 841 (1990).

\bibitem{pumplin} J. Pumplin, Phys. Rev. {\bf D41}, 900 (1990).

\bibitem{pr91} P. A. R\c{a}czka, Phys. Rev., {\bf D43}, R9 (1991).

\bibitem{chylkat} J. Chyla, A. Kataev and S. A. Larin, Phys. Lett.
 {\bf B267}, 269 (1991).

\bibitem{haru92} M. Haruyama, Phys. Rev. {\bf D45}, 930 (1992).

\bibitem{pich92b} A. Pich, Proceedings of the Second Workshop on
 Tau Lepton Physics, Columbus, Ohio, 8-11 Sept. 1992.

\bibitem{narpi93} S. Narison and A. Pich, Phys. Lett. {\bf B304},
 359 (1993).

\bibitem{aleph} ALEPH Collab., D. Buskulic et al.,
 Phys. Lett. {\bf B307}, 209 (1993).

\bibitem{chetyrkin} K. G. Chetyrkin, Phys. Lett. {\bf B307},
 169 (1993).

\bibitem{chetkwiat} K. G. Chetyrkin and A. Kwiatkowski, Z. Phys.
 {\bf C59}, 525 (1993).

\bibitem{truong} T. N. Truong, Phys. Rev. {\bf D47}, 3999 (1993).

\bibitem{mattingly} A. C. Mattingly, preprint DE-FG05-92ER40717-8.

\bibitem{brodsky} S. J. Brodsky and Hung Jung Lu,
 Phys.\ Rev.\ {\bf D51}, 3652 (1995).

\bibitem{katstar} A. L. Kataev and V. V. Starshenko, preprint
 CERN-TH.7400/94 (hep-ph/9408395), also Mod.\ Phys.\ Lett.\
 {\bf A10}, 235 (1995).

\bibitem{pr95} The results of Ref.1 are of course confirmed when the
 concrete form (6) of the condition on the ``natural'' schemes
 is used.

\bibitem{rge} C. J. Maxwell, Phys. Rev. {\bf D28}, 2037 (1983).

\bibitem{pdg} Review of Particle Properties, Particle Data
 Group, L. Montanet et al., Phys. Rev. {\bf D50}, 1173 (1994).

\bibitem{marciano} W. J. Marciano, Phys.\ Rev.\ {\bf D29}, 580 (1984).
See W. Bernreuther, preprint PITHA 94/31 (hep-ph/9409390),
 for discussion of
the accuracy of this formula. We have used $m_{c}=1.55\,\mbox{GeV}$,
$m_{b}=4.73\,\mbox{GeV}$.

\bibitem{pich94} A. Pich, hep-ph/9412273 (talk given at the QCD~94
 Workshop).

\bibitem{narison94} S. Narison, preprint CERN-TH-7506-94
 (hep-ph/9412295).


\end{thebibliography}
\end{document}